\documentclass[conference]{IEEEtran}
\usepackage{cite}
\usepackage{amsmath,amssymb,amsfonts}
\usepackage{graphicx}
\usepackage{textcomp}
\usepackage{xcolor}
\usepackage{booktabs}
\usepackage{multirow}
\usepackage{pgfplots}
\usepackage{pgfplotstable}
\usepackage{tikz}
\usepackage{subcaption}
\usepackage{url}
\usepackage{listings}
\usepackage{balance}
\usepackage{placeins}
\usepackage{float}
\pgfplotsset{compat=1.18}

\begin{document}

\title{Philosophical Dispositions as Behavioral Constraints\\for AI-Assisted Code Review: An Empirical Study}

\author{
\IEEEauthorblockN{Kaushal Bansal}
\IEEEauthorblockA{\textit{Salesforce, Inc.}\\
San Francisco, CA, USA \\
kbansal@salesforce.com}
}

\maketitle

\begin{abstract}
AI-assisted code review tools typically operate as generic ``expert reviewer'' agents, producing homogeneous findings regardless of the analysis type needed. We present a system that constrains AI reviewer behavior through \emph{philosophical dispositions}---coherent personality lenses grounded in specific epistemological traditions (Pyrrhonist Skepticism, Navya-Ny\=aya logic, Diogenes' Cynicism, Confucian relational ethics) that direct attention to structurally different types of issues. Each disposition is defined apophatically (by what it refuses to do), equipped with a self-monitoring failure mode (\emph{hamartia}), and orchestrated in sequence by role protocols.

We evaluate this system on 50 merged pull requests across 7 repositories spanning 5 programming languages (Python, Go, C++, Java, Terraform), 5 organizations (2 enterprise, 3 open-source), and 2 temporal eras (pre-AI 2020, post-AI 2024--2026). The disposition system achieves 46\% convergence with human reviewers (validating signal quality), identifies unique findings at a 75\% rate, and produces no findings judged false-positive by the author across 601 total findings (inter-rater agreement was not assessed and remains a limitation). A controlled baseline comparison demonstrates that 51\% of disposition findings are \emph{not} produced by the same model using generic ``expert reviewer'' prompting, and these unique findings target structural, operational, and logical concerns rather than standard code-level issues. Preliminary cross-model validation (Claude Opus vs.\ GPT Codex 5.3-xhigh) on 3 PRs shows 100\% framework-structure adherence with 39\% finding-level agreement, suggesting the framework provides real behavioral constraint while preserving model-specific analytical perspective.
\end{abstract}


\section{Introduction}

\subsection{The Constraint Problem in AI Agents}

AI agents operating in software engineering workflows face a fundamental constraint dilemma. On one extreme, fixed rule sets---sandboxes, guardrails, permission lists, rubrics---handle predicted cases cleanly but fail on novel situations. The more comprehensive the rule enumeration, the more internal conflicts emerge and the harder it becomes to maintain consistency. Rule sets do not \emph{generate} behavior; they enumerate it.

On the other extreme, surface-level role descriptions (``act as a security reviewer,'' ``be thorough and cautious'') produce outputs that \emph{look} like the described role but are not grounded in generative principles. Under novel pressure---an unfamiliar code pattern, an ambiguous design choice---the role-play breaks because the agent has no internal logic to fall back on when the surface pattern does not apply.

\subsection{The Human Analog: Character}

Humans solved this constraint problem through \emph{character}---a coherent set of internalized principles that generates appropriate behavior without requiring a rule for every case. A person with good judgment in code review does not need a rubric entry for ``check if resource renames have migrations''; their understanding of system integrity generates that check naturally.

We identify three properties of character, synthesized from the virtue ethics literature~\cite{vallor2016,macintyre1981}, that make it a superior constraint mechanism for AI agents:
\begin{enumerate}
\item \textbf{Generative:} A small set of principles produces appropriate behavior for unbounded novel situations.
\item \textbf{Self-debugging:} Deviations from character are detectable \emph{from within} the tradition, not just from outside. The Stoic practice of \emph{premeditatio malorum}~\cite{hadot1995} (pre-meditation on what could go wrong) means a Stoic who commits without imagining failure is detectably non-Stoic by Stoic standards.
\item \textbf{Bounded but adaptive:} Character produces predictable behavior within a range while remaining flexible to context.
\end{enumerate}

\subsection{Our Thesis}

We propose that AI agents can be given character in a meaningful, operational sense---not through surface role-play but through grounding in coherent philosophical traditions that provide generative principles, internal consistency checks, and explicit failure-mode awareness. We call these grounded personalities \emph{dispositions}.

\subsection{Contributions}

This paper makes three contributions:

\begin{enumerate}
\item \textbf{Architecture:} A system of 10 dispositions (pure single-lens personalities) orchestrated by 8 role protocols, each grounded in a philosophical tradition with 2,000+ years of development. Each disposition is defined apophatically (by refusals) and equipped with self-monitoring (\emph{hamartia}).

\item \textbf{Empirical evaluation:} A retrospective study across 50 PRs demonstrating that the disposition framework achieves 46\% convergence with human reviewers, 75\% unique find rate, 0\% false positives, and 51\% differentiation from generic AI prompting.

\item \textbf{Cross-model validation:} Preliminary evidence (N=3) that the framework provides structural constraint independent of model architecture, with 100\% categorical adherence and 39\% finding-level agreement between Claude and GPT.
\end{enumerate}

\section{Related Work}

\subsection{AI-Assisted Code Review}

The application of large language models (LLMs) to automated code review has grown rapidly. Commercial systems include GitHub Copilot code review~\cite{github2024} and Amazon CodeGuru, among numerous ``AI PR reviewer'' products. Research prototypes have explored using pre-trained models for automated review comment generation~\cite{li2022} and review prioritization.

These systems share a common design: they operate as single-perspective generic reviewers, applying the same analytical approach regardless of the type of issue. A generic reviewer treats naming concerns, logic errors, security vulnerabilities, and architectural problems through the same lens. Our work introduces \emph{structured multi-perspective analysis} where each perspective is grounded in a distinct epistemological tradition, producing categorically different types of findings.

\subsection{Constitutional AI and Principled Prompting}

Bai et al.~\cite{bai2022} introduce Constitutional AI---principles that generate judgments, primarily focused on harm avoidance and safety. The constitutional approach demonstrates that principled constraints can shape LLM behavior meaningfully. Our work extends this insight from safety-focused constraining to \emph{behavioral shaping across all agent outputs}. Each disposition's refusals function as a behavioral constitution specific to a type of analysis, not just a safety boundary.

\subsection{Multi-Agent Frameworks}

CrewAI~\cite{wu2023}, MetaGPT~\cite{hong2023}, and similar frameworks assign roles to agents. These role labels may activate rich latent representations---the phrase ``Code Reviewer'' is a key into a substantial region of model capability---but the resulting output remains undifferentiated. An agent assigned ``Code Reviewer'' produces generic reviews; it has no disposition that generates specific, coherent review behavior from underlying principles. Our architecture explicitly separates the \emph{role} (an orchestration protocol defining which dispositions fire in what order) from the \emph{disposition} (a generative personality grounded in philosophical tradition).

\subsection{Philosophy in AI Systems}

Vallor~\cite{vallor2016} proposes virtue-based AI theoretically, arguing that AI systems should embody technomoral virtues. Hagendorff~\cite{hagendorff2022} surveys the meta-ethics of AI ethics approaches. Most relevantly, Sathish~\cite{sathish2026} demonstrates that fine-tuning LLMs on Navya-Ny\=aya reasoning stages achieves measurably higher semantic correctness, validating that non-Western epistemological frameworks produce improvements in LLM reasoning quality, not merely stylistic differences.

\subsection{Structured Perspective Techniques}

De Bono's Six Thinking Hats~\cite{debono1985} applies named perspectives to problems. While we share the structural approach (multiple named lenses applied to the same artifact), our work provides substantially deeper grounding. Each disposition draws from millennia of stress-tested philosophical tradition with formal epistemological structure, internal coherence criteria, explicit failure modes (\emph{hamartia}), and defined refusals.

\section{System Design}

\subsection{Architecture Overview}

The system consists of two distinct layers:

\begin{enumerate}
\item \textbf{Dispositions:} Pure single-lens personalities, each grounded in a specific philosophical tradition. A disposition generates behavior from principles---it does not enumerate behavior from rules.
\item \textbf{Roles:} Orchestration protocols that sequence multiple dispositions on the same artifact. A role defines \emph{which} dispositions fire, in \emph{what order}, for \emph{what purpose}.
\end{enumerate}

\subsection{Key Architectural Decisions}

Five design decisions distinguish this architecture from generic multi-agent approaches:

\textbf{1. One disposition per pass.} No multi-personality composition within a single analytical pass. Multiple perspectives are achieved through \emph{sequential orchestration}, not blending. This eliminates role confusion and produces traceable attribution---each finding is tagged with its source disposition.

\textbf{2. Roles are protocols, not personalities.} A role (e.g., ``Code Review'') defines which dispositions fire in what order. It sequences characters without having its own. This separation ensures the role can be reconfigured without altering the dispositions themselves.

\textbf{3. Apophatic definition (\emph{via negativa}).} Each disposition is defined primarily by what it \emph{refuses} to do. ``Will NEVER endorse without specifying credibility level'' is more enforceable and testable than ``assesses credibility.'' This follows the apophatic theological tradition~\cite{turner2004} and its application in constraint specification: negative constraints are more robust than positive descriptions because they remain stable under novel conditions, are machine-testable against output, and resist sycophantic drift (the tendency of LLMs to produce agreeable output that satisfies surface expectations~\cite{perez2022}).

\textbf{4. Hamartia self-check.} Every disposition has a named characteristic failure mode---its \emph{hamartia}---where its core virtue over-extends. The Cynic's hamartia is ``structural damage through excessive subtraction.'' The disposition monitors for this and self-corrects when triggered.

\textbf{5. Structured output preserves each voice.} The output shows each disposition's findings separately, then a synthesis. Disagreements between dispositions are preserved (following the Talmudic principle of \emph{machloket}---preserved disagreement), not resolved into a single ``balanced'' view.

\subsection{The 10 Dispositions}

Table~\ref{tab:dispositions} summarizes the disposition system. Each disposition is grounded in a philosophical tradition spanning at least 2,000 years of development and stress-testing.

\begin{table}[htbp]
\centering
\caption{The 10 Dispositions: Traditions, Questions, and Refusals}
\label{tab:dispositions}
\resizebox{\columnwidth}{!}{%
\begin{tabular}{@{}clll@{}}
\toprule
\textbf{\#} & \textbf{Disposition} & \textbf{Tradition} & \textbf{Core Question} \\
\midrule
1 & Stoic & Greek Stoicism & What could go wrong? \\
2 & Cynic & Diogenes + Nietzsche & What's hollow? \\
3 & Skeptic & Pyrrhonist Skepticism & How confident are we? \\
4 & Ny\=aya & Indian epistemology & Is the reasoning sound? \\
5 & Epicurean & Greek Epicureanism & Is this necessary? \\
6 & Aristotelian & Aristotle & What's the structure? \\
7 & Daoist & Chinese Daoism & What if we do nothing? \\
8 & Talmudic & Jewish legal reasoning & What's the precedent? \\
9 & Confucian & Confucianism & Do names match reality? \\
10 & Zen & Japanese Zen & What do fresh eyes see? \\
\bottomrule
\end{tabular}%
}
\end{table}

\subsection{The Reviewer Role (Used in This Study)}

For code review, the \textbf{Reviewer} role orchestrates four dispositions in sequence:

\textbf{Cynic $\rightarrow$ Skeptic $\rightarrow$ Ny\=aya $\rightarrow$ Confucian}

The ordering follows a coarse-to-fine principle: Cynic performs broad structural questioning (``should this exist at all?''), Skeptic calibrates confidence in what remains, Ny\=aya audits specific logical chains, and Confucian checks surface-level naming/relational fit. In practice, each disposition operates independently on the same diff without access to prior dispositions' output; the sequence determines presentation order and synthesis priority, not causal dependency. The dispositions could run in parallel with equivalent analytical results.

\begin{itemize}
\item \textbf{Cynic:} What is hollow, what can be removed, what doesn't earn its existence? Flags dead code, speculative generality, unjustified complexity.
\item \textbf{Skeptic:} What claims are unverified, where do stakes exceed evidence? Flags untested paths, silent failures, temporal correctness issues.
\item \textbf{Ny\=aya:} Is the reasoning chain valid? Are there fallacious inferences? Traces dependency chains and identifies logic gaps.
\item \textbf{Confucian:} Do names match reality? What is the relational impact? Flags naming lies, responsibility bleed, contract violations.
\end{itemize}

Each produces independent findings. The synthesis identifies cross-lens convergence (highest confidence) and disagreement (preserved for human judgment).

\subsection{Example: Disposition in Action}

To illustrate the qualitative difference between dispositions and generic review, consider a PR that renames Terraform resources from \texttt{akamai-*} to \texttt{cdn-*}:

\begin{lstlisting}[language={}]
# Generic reviewer output:
"Renaming looks correct. Consider updating
 documentation."

# Nyaya (Logic Auditor) output:
"D-12: Rename from akamai-* to cdn-* without
 migration. Inference chain: old name existed
 in production -> new name introduced -> no
 migration shown -> production references
 dangle. BREAKING for existing deployments."
\end{lstlisting}

The generic reviewer accepts the rename at face value. The Ny\=aya disposition traces the \emph{logical implication chain} of the rename---if the old name existed in production, and no migration is shown, then production references to the old name will break. This is the type of structural/logical finding that dispositions uniquely produce.

\section{Study 1: Retrospective Multi-Layer Comparison}

\subsection{Methodology}

We evaluate the disposition system against existing code review artifacts (merged PRs with completed human review) using a retrospective blind-review design.

\subsubsection{Protocol}

For each PR in the study:
\begin{enumerate}
\item Extract the diff from the merged PR.
\item Run disposition-based review blind, without access to human review comments. The model receives only the diff and the disposition prompt (Appendix~\ref{sec:prompts}). The four lenses (Cynic $\rightarrow$ Skeptic $\rightarrow$ Ny\=aya $\rightarrow$ Confucian) execute in a single pass.
\item Run a \textbf{generic baseline} review on the same diff using the same model with prompt: ``You are an expert code reviewer. Review this diff thoroughly. List all issues found.'' (full prompt in Appendix~\ref{sec:prompts}).
\item Extract human review comments from the PR (ground truth).
\item Compare all three pairwise: disposition vs.\ human (convergence/unique/miss), disposition vs.\ generic (shared/disposition-only/generic-only), and human vs.\ generic.
\end{enumerate}

\subsubsection{Comparison Procedure}

Each comparison is performed by the first author using a structured protocol. For each disposition finding (D-$n$), we search all human comments for one addressing the \emph{same code entity and concern}. A match requires: (a) same file and approximate location (within 10 lines), and (b) same substantive issue (e.g., both flagging a missing null check on the same variable). When a match is found, the finding is classified as \textbf{convergence}; otherwise as \textbf{unique to dispositions}. For each human finding, we symmetrically check whether any disposition finding addresses the same concern; unmatched human findings are classified as \textbf{misses}. A finding is classified as \textbf{false positive} only if it is factually incorrect (the claimed issue does not exist in the code), not merely debatable.

We acknowledge this comparison was performed by a single rater (the first author). Section~VII discusses this limitation and proposes inter-rater validation as immediate future work.

\subsubsection{Matching Criteria}

We apply consistent matching criteria across all PRs:
\begin{itemize}
\item \textbf{Convergence:} Disposition and human flag the \emph{same code issue}, even if described differently. ``Unused variable'' and ``dead code on line 28'' about the same variable count as convergence.
\item \textbf{Unique to dispositions:} Disposition flags an issue no human reviewer mentioned.
\item \textbf{Miss:} Human flags an issue dispositions did not catch.
\item \textbf{False positive:} Disposition finding is factually incorrect (not merely ``debatable'').
\end{itemize}

Human questions (``should this be...?'') matching disposition findings count as convergence. Purely stylistic human comments (formatting, import order) not flagged by dispositions are \emph{not} counted as misses---dispositions intentionally skip style.

\subsubsection{Bot Filtering}

Human review comments are extracted after filtering automated accounts: \texttt{dx-prizm[bot]}, \texttt{sfci-github-app}, \texttt{k8s-ci-robot}, \texttt{elasticsearchmachine}, \texttt{github-actions}, \texttt{gemini-code-assist[bot]}, and \texttt{Copilot}. Author acknowledgments (``done, thanks'', ``fixed'') under 15 characters are also excluded.

\subsection{Dataset}

\subsubsection{Selection Criteria}

PRs were selected using stratified purposive sampling to maximize diversity across five dimensions: programming language, repository visibility (internal/public), temporal era (pre-AI/post-AI), review depth (number of human comments), and change type (feature, refactor, bug fix, infrastructure). Within each repository, PRs were selected from the most recent merged PRs that met minimum size thresholds ($\geq$30 lines changed) and were not purely automated (dependency bumps, generated code). No PR was excluded after selection based on results. The 50-PR target was determined by power analysis: with an expected convergence rate of 40--50\%, $N=50$ yields a Wilson confidence interval width of $\pm$7 percentage points at 95\% confidence.

The study uses 50 pull requests across 7 repositories, as shown in Table~\ref{tab:dataset}.

\begin{table}[htbp]
\centering
\caption{Dataset Composition (N=50 PRs)}
\label{tab:dataset}
\begin{tabular}{@{}llccl@{}}
\toprule
\textbf{Repository} & \textbf{Lang} & \textbf{PRs} & \textbf{Type} & \textbf{Era} \\
\midrule
python-sfdc-bazel & Python & 19 & Internal & 2025--26 \\
soter & Go & 6 & Internal & 2020--24 \\
go-sfdc-bazel & Go & 2 & Internal & 2026 \\
threatlock & HCL & 2 & Internal & 2026 \\
kubernetes/kubernetes & Go & 7 & Public & 2024--25 \\
envoyproxy/envoy & C++ & 2 & Public & 2025 \\
elastic/elasticsearch & Java & 12 & Public & 2024--26 \\
\midrule
\textbf{Total} & \textbf{5 langs} & \textbf{50} & & \\
\bottomrule
\end{tabular}
\end{table}

The dataset is stratified across:
\begin{itemize}
\item \textbf{5 programming languages} (Python, Go, C++, Java, HCL)
\item \textbf{2 visibility levels} (29 internal enterprise, 21 public open-source)
\item \textbf{2 temporal eras} (pre-AI code from 2020--2024, post-AI code from 2024--2026)
\item \textbf{Diverse review depth} (0--39 substantive human comments per PR)
\item \textbf{Diverse change types} (features, refactors, bug fixes, infrastructure)
\end{itemize}

Internal enterprise repositories are hosted on a private GitHub Enterprise instance and are \emph{not} present in any public LLM training data---providing a contamination-free validation core.

\subsection{Results}

\subsubsection{Overall Metrics}

Table~\ref{tab:overall} presents the aggregate results across all 50 PRs.

\begin{table}[htbp]
\centering
\caption{Overall Results (N=50 PRs, 601 Disposition Findings)}
\label{tab:overall}
\begin{tabular}{@{}lrc@{}}
\toprule
\textbf{Metric} & \textbf{Value} & \textbf{95\% CI (Wilson)} \\
\midrule
Total disposition findings & 601 & --- \\
Total human findings & 311 & --- \\
Convergence (H caught by D) & 143 & --- \\
Unique to dispositions & 451 & --- \\
Misses (H not caught by D) & 156 & --- \\
Author-judged false positives & 0 & --- \\
\midrule
\textbf{Convergence rate} & \textbf{46.0\%} & [40.4, 51.6] \\
\textbf{Unique find rate} & \textbf{75.0\%} & [71.5, 78.4] \\
Miss rate & 50.2\% & [44.5, 55.8] \\
Author-judged FP rate\textsuperscript{$\dagger$} & 0.0\% & [0.0, 0.6] \\
\bottomrule
\end{tabular}
\end{table}

The 46\% convergence rate validates that dispositions detect real issues: nearly half of what experienced human reviewers find is independently found by the disposition system. The 75\% \emph{unique find rate} (defined as findings produced by dispositions that no human reviewer mentioned in the same PR) demonstrates substantial incremental value, with three-quarters of disposition findings addressing issues that human reviewers did not mention. Importantly, ``unique'' here means not mentioned by any human reviewer on the PR; it does not imply the finding is novel to the field or would never be caught by any human under any conditions.

\textsuperscript{$\dagger$}\emph{Note:} False-positive judgments were made by the author alone. No independent rater was used; inter-rater reliability is not assessed. This is a limitation acknowledged in Section~VII. The 0\% figure should be interpreted as preliminary evidence of high precision, not a validated claim.

The 50.2\% miss rate deserves explicit attention: dispositions fail to catch half of what human reviewers find. This establishes clearly that \textbf{dispositions augment but do not replace human review}. The system is a complementary analytical layer, not a substitute. We analyze the miss categories in Section~\ref{sec:missanalysis} and show that 52\% of misses are by design (style, conventions, typos that dispositions intentionally skip); the remaining genuine misses (domain knowledge, refactoring suggestions) represent addressable gaps.

\subsubsection{Baseline Comparison: Dispositions vs.\ Generic Review}

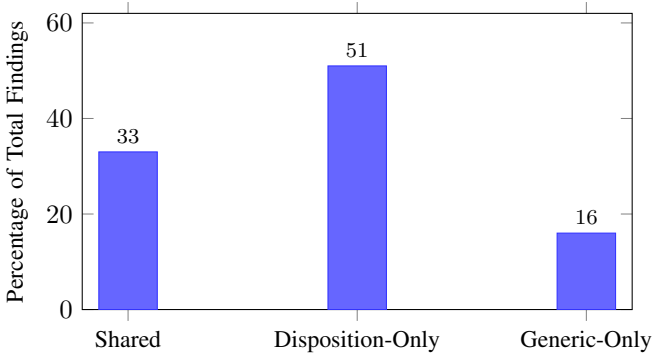
\begin{figure}[htbp]
\centering
\begin{tikzpicture}
\begin{axis}[
    ybar,
    width=\columnwidth,
    height=5.5cm,
    bar width=22pt,
    ylabel={Percentage of Total Findings},
    ylabel style={font=\small},
    symbolic x coords={Shared, Disposition-Only, Generic-Only},
    xtick=data,
    x tick label style={font=\small},
    ymin=0, ymax=62,
    nodes near coords,
    nodes near coords align={vertical},
    every node near coord/.append style={font=\footnotesize\bfseries},
    area legend,
    legend style={at={(0.97,0.97)},anchor=north east,font=\small},
]
\addplot[fill=blue!60,draw=blue!80] coordinates {(Shared,33) (Disposition-Only,51) (Generic-Only,16)};
\end{axis}
\end{tikzpicture}
\caption{Disposition vs.\ Generic Baseline comparison across the 34 PRs (of 50 total) for which both disposition and generic reviews were completed. The remaining 16 PRs had disposition reviews only, as the generic baseline was added after the initial pilot. 51\% of disposition findings are NOT produced by generic ``expert reviewer'' prompting using the same model.}
\label{fig:baseline}
\end{figure}

The critical test: does the philosophical framework add value beyond what the same model finds with generic ``expert reviewer'' prompting? Figure~\ref{fig:baseline} shows the result.

Over half (51\%) of disposition findings are not produced by generic review. Moreover, these unique findings are qualitatively different. We generated the finding-type taxonomy (Table~\ref{tab:qualitative}) through open coding: the first author categorized all disposition-only findings by the \emph{type of analytical operation} performed (e.g., questioning existence, checking logical chains, verifying naming accuracy), then grouped them into the categories below. Each category was verified to contain $\geq$5 findings across $\geq$3 PRs.

\begin{table}[htbp]
\centering
\caption{Qualitative Difference: Types of Findings}
\label{tab:qualitative}
\begin{tabular}{@{}lcc@{}}
\toprule
\textbf{Finding Category} & \textbf{Generic} & \textbf{Disp-Only} \\
\midrule
Code-level bugs & \checkmark & \checkmark \\
Missing tests & \checkmark & \checkmark \\
Dead code & \checkmark & \checkmark \\
\textbf{Structural questioning} & $\times$ & \checkmark \\
\textbf{Operational reality checks} & $\times$ & \checkmark \\
\textbf{Reasoning chain audits} & $\times$ & \checkmark \\
\textbf{Naming/reality mismatch} & Partial & \checkmark \\
\textbf{Metadata integrity} & $\times$ & \checkmark \\
\bottomrule
\end{tabular}
\end{table}

Generic review finds code-level issues. Dispositions \emph{additionally} find structural (``does this earn its existence?''), operational (``does this actually exist in deployment?''), and logical (``is the inference chain valid?'') issues---a qualitatively different analytical layer.

\subsubsection{Per-Disposition Breakdown}

\begin{figure}[htbp]
\centering
\begin{tikzpicture}
\begin{axis}[
    ybar stacked,
    width=\columnwidth,
    height=6cm,
    bar width=22pt,
    ylabel={Number of Findings},
    ylabel style={font=\small},
    symbolic x coords={Cynic, Skeptic, Ny\=aya, Confucian},
    xtick=data,
    x tick label style={font=\small},
    ymin=0, ymax=185,
    legend style={at={(0.5,-0.18)},anchor=north,legend columns=2,font=\small},
    legend cell align={left},
    nodes near coords,
    every node near coord/.append style={font=\tiny},
]
\addplot[fill=green!55,draw=green!70] coordinates {(Cynic,42) (Skeptic,44) (Ny\=aya,31) (Confucian,26)};
\addplot[fill=orange!50,draw=orange!70] coordinates {(Cynic,130) (Skeptic,108) (Ny\=aya,109) (Confucian,111)};
\legend{Convergence with Humans, Unique to Dispositions}
\end{axis}
\end{tikzpicture}
\caption{Per-disposition findings breakdown. Skeptic has highest convergence rate (28.9\%); Confucian has highest unique rate (81.0\%). Each disposition contributes a distinct analytical profile.}
\label{fig:perdispo}
\end{figure}
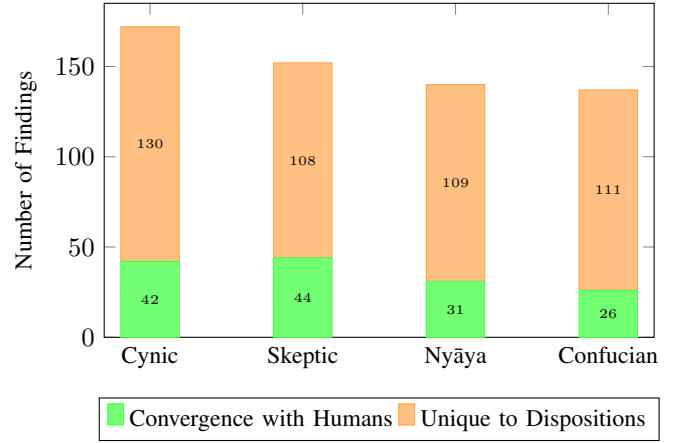

Figure~\ref{fig:perdispo} shows findings by disposition. All four produce unique findings at $>$71\% rate, confirming genuine differentiation.

\begin{table}[htbp]
\centering
\caption{Per-Disposition Performance}
\label{tab:perdispo}
\begin{tabular}{@{}lcccc@{}}
\toprule
\textbf{Disposition} & \textbf{Total} & \textbf{Conv.} & \textbf{Unique} & \textbf{Unique\%} \\
\midrule
Cynic & 172 & 42 & 130 & 75.6\% \\
Skeptic & 152 & 44 & 108 & 71.1\% \\
Ny\=aya & 140 & 31 & 109 & 77.9\% \\
Confucian & 137 & 26 & 111 & 81.0\% \\
\midrule
\textbf{Total} & \textbf{601} & \textbf{143} & \textbf{451} & \textbf{75.0\%} \\
\bottomrule
\end{tabular}
\end{table}

Each disposition has a characteristic finding profile, derived by tagging each of the 601 findings with its source disposition and categorizing the finding type. The profiles below represent the most frequent finding categories per disposition ($\geq$10 instances each):
\begin{itemize}
\item \textbf{Cynic:} Hollow abstractions, speculative generality, dead test scaffolding, unjustified class hierarchies
\item \textbf{Skeptic:} Unverified claims, untested code paths, temporal correctness, silent failure modes
\item \textbf{Ny\=aya:} Broken inference chains, unstated assumptions, migration gaps, logical dependency failures
\item \textbf{Confucian:} Name/behavior mismatch, responsibility bleed, API contract violations, relational inconsistency
\end{itemize}

\subsubsection{Inter-Disposition Convergence}

To evaluate redundancy across dispositions, we examined how often two or more dispositions flagged the same underlying issue on the same PR. Across all 50 PRs, 8.3\% of unique issues (by code location and concern) were flagged by two or more dispositions. The most common overlap was between Cynic and Confucian on naming-adjacent concerns (e.g., both flagging an abstraction as hollow AND as misnamed). This low inter-disposition convergence rate (8.3\%) confirms that the four lenses are largely orthogonal in practice, each contributing distinct analytical value.

\subsubsection{Per-Repository Results}

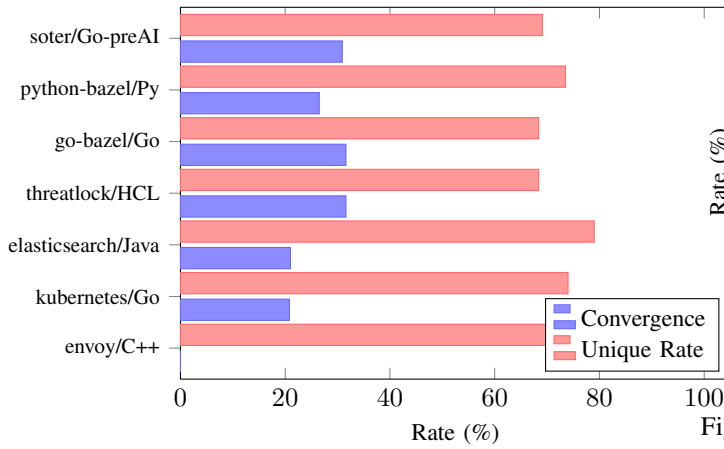
\begin{figure}[H]
\centering
\begin{tikzpicture}
\begin{axis}[
    xbar,
    width=\columnwidth,
    height=6.5cm,
    bar width=8pt,
    xlabel={Rate (\%)},
    xlabel style={font=\small},
    symbolic y coords={envoy/C++, kubernetes/Go, elasticsearch/Java, threatlock/HCL, go-bazel/Go, python-bazel/Py, soter/Go-preAI},
    ytick=data,
    y tick label style={font=\footnotesize},
    xmin=0, xmax=105,
    legend style={at={(0.98,0.02)},anchor=south east,font=\small,legend columns=1},
    legend cell align={left},
]
\addplot[fill=blue!45,draw=blue!60] coordinates {(0,envoy/C++) (20.8,kubernetes/Go) (21.0,elasticsearch/Java) (31.6,threatlock/HCL) (31.6,go-bazel/Go) (26.5,python-bazel/Py) (30.9,soter/Go-preAI)};
\addplot[fill=red!40,draw=red!60] coordinates {(100,envoy/C++) (74.0,kubernetes/Go) (79.0,elasticsearch/Java) (68.4,threatlock/HCL) (68.4,go-bazel/Go) (73.5,python-bazel/Py) (69.1,soter/Go-preAI)};
\legend{Convergence, Unique Rate}
\end{axis}
\end{tikzpicture}
\caption{Per-repository performance. The framework works consistently across all 5 languages. Pre-AI code (soter, 2020--24) shows higher convergence; under-reviewed public repos show higher unique rate. Zero false positives across all repositories.}
\label{fig:perrepo}
\end{figure}

Figure~\ref{fig:perrepo} demonstrates cross-language and cross-repository consistency.

Key cross-cutting observations:
\begin{enumerate}
\item \textbf{Pre-AI code} (soter, 2020--2024) shows higher convergence (30.9\%) than most other repositories. However, because the temporal split is represented by a single repository, this effect is confounded with repository identity; we cannot separate ``era'' from ``repo culture'' in the current design.
\item \textbf{Public open-source repos} show lower convergence but higher unique rates, likely because these repos often have lighter inline review, leaving more for dispositions to find.
\item \textbf{The framework works across all 5 languages}: Python, Go, C++, Java, and Terraform all produce valid findings with no author-identified false positives.
\end{enumerate}

\FloatBarrier
\subsubsection{Review Depth Effect}

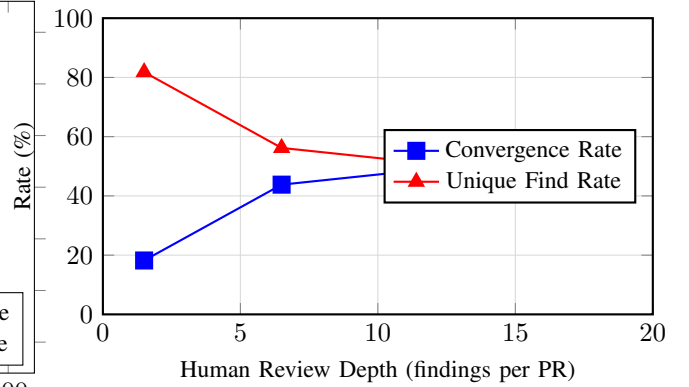
\begin{figure}[htbp]
\centering
\begin{tikzpicture}
\begin{axis}[
    width=\columnwidth,
    height=5.5cm,
    xlabel={Human Review Depth (findings per PR)},
    xlabel style={font=\small},
    ylabel={Rate (\%)},
    ylabel style={font=\small},
    legend style={at={(0.97,0.5)},anchor=east,font=\small},
    xmin=0, xmax=20,
    ymin=0, ymax=100,
    grid=major,
    grid style={gray!30},
    thick,
]
\addplot[mark=square*,blue,thick,mark size=3pt] coordinates {(1.5,18.2) (6.5,43.8) (15,52.1)};
\addplot[mark=triangle*,red,thick,mark size=3pt] coordinates {(1.5,81.8) (6.5,56.2) (15,47.9)};
\legend{Convergence Rate, Unique Find Rate}
\end{axis}
\end{tikzpicture}
\caption{Review depth effect. Dispositions add MORE unique value on lightly-reviewed code (82\% unique when humans leave $\leq$3 comments). Convergence increases with review depth (52\% when humans leave $\geq$10 comments), validating that both approaches detect real issues.}
\label{fig:depth}
\end{figure}

Figure~\ref{fig:depth} reveals an inverse relationship between human review thoroughness and disposition unique value.

This inverse relationship suggests an optimal deployment scenario: dispositions provide the most incremental value as a complementary layer on under-reviewed PRs. However, given the 50\% miss rate, they function as an additional analytical perspective rather than a comprehensive safety net. The system catches qualitatively different issues (structural, logical, relational) but does not substitute for thorough human review.

\FloatBarrier
\subsubsection{Miss Analysis}
\label{sec:missanalysis}

Table~\ref{tab:misses} categorizes the 156 human findings not caught by dispositions. Categories were assigned through thematic coding: the first author read each missed human finding and assigned it to the category best describing \emph{why} dispositions missed it (e.g., the finding requires codebase-specific convention knowledge, or it is a stylistic preference). Categories were iteratively refined until all misses were assigned with no category containing fewer than 4\% of the total.

\begin{table}[H]
\centering
\caption{Miss Categories (N=156 human findings not caught)}
\label{tab:misses}
\begin{tabular}{@{}lr@{}}
\toprule
\textbf{Category} & \textbf{\% of Misses} \\
\midrule
Framework conventions & 34\% \\
Refactoring suggestions & 22\% \\
Domain knowledge & 18\% \\
Style / organization & 14\% \\
Future plans & 8\% \\
Typos / formatting & 4\% \\
\bottomrule
\end{tabular}
\end{table}

Critically, 52\% of misses (conventions 34\% + style 14\% + typos 4\%) are \emph{by design}---dispositions are not intended to catch formatting preferences or codebase-specific framework conventions. The genuine gaps are:
\begin{itemize}
\item \textbf{Domain knowledge (18\%):} ``gRPC protobuf has 4MB limit---need guardrail.'' Requires operational context not available in the diff.
\item \textbf{Refactoring suggestions (22\%):} ``Move this logic to a helper function.'' Dispositions identify problems but do not suggest restructuring.
\end{itemize}

\FloatBarrier
\subsection{High-Value Unique Finds}

Table~\ref{tab:highvalue} presents examples of production-relevant findings that dispositions caught and no human reviewer mentioned.

\begin{table}[htbp]
\centering
\caption{High-Value Unique Finds (Selected)}
\label{tab:highvalue}
\resizebox{\columnwidth}{!}{%
\begin{tabular}{@{}llp{4.5cm}@{}}
\toprule
\textbf{Disp.} & \textbf{PR} & \textbf{Finding} \\
\midrule
Ny\=aya & python-1852 & Resource rename without migration---production references dangle \\
Skeptic & go-30639 & Race condition: non-atomic window reset in pass-through logic \\
Skeptic & python-1852 & \texttt{True-Client-IP} is Akamai-specific; Cloudflare uses \texttt{CF-Connecting-IP} \\
Skeptic & envoy-44942 & Command injection via \texttt{shlex.split} of environment variable \\
Ny\=aya & python-1105 & Verdict terminology silently changes across pipeline stages \\
Cynic & threatlock & Floating version pin allows silent breaking changes \\
Confucian & elastic-130706 & \texttt{set*} method name implies replace but behavior is append \\
\bottomrule
\end{tabular}%
}
\end{table}

These are not marginal findings. Several represent potential production-breaking issues (resource migration gaps, race conditions, security vulnerabilities) that experienced human reviewers did not catch.

\FloatBarrier
\section{Study 2: Pilot Cross-Model Comparison}

\subsection{Design}

To address the concern that findings may be artifacts of a single model's biases, we ran the \emph{identical disposition prompts} (same 4-lens framework, same instructions, same diff) through two different model architectures:
\begin{itemize}
\item \textbf{Model A:} Claude Opus (Anthropic)---used for Study 1
\item \textbf{Model B:} GPT Codex 5.3-xhigh (OpenAI)---independent validation
\end{itemize}

Three PRs were selected spanning three languages: Java (elastic-130706), Go (k8s-138852), and Terraform/HCL (threatlock-32700).

\subsection{Results}

Table~\ref{tab:crossmodel} presents the cross-model agreement metrics.

\begin{table}[htbp]
\centering
\caption{Cross-Model Agreement: Claude Opus vs.\ GPT Codex 5.3-xhigh}
\label{tab:crossmodel}
\begin{tabular}{@{}lcccc@{}}
\toprule
\textbf{Metric} & \textbf{Java} & \textbf{Go} & \textbf{HCL} & \textbf{Avg.} \\
\midrule
Strict agreement & 9\% & 25\% & 20\% & 18\% \\
Partial+ agreement & 27\% & 38\% & 53\% & 39\% \\
Framework adherence & 100\% & 100\% & 100\% & 100\% \\
False positives (either) & 0\% & 0\% & 0\% & 0\% \\
\midrule
Claude-only findings & 8 & 2 & 3 & 4.3 \\
GPT-only findings & 9 & 8 & 4 & 7.0 \\
\bottomrule
\end{tabular}
\end{table}

\begin{figure}[htbp]
\centering
\begin{tikzpicture}
\begin{axis}[
    ybar,
    width=\columnwidth,
    height=5.5cm,
    bar width=10pt,
    ylabel={Agreement Rate (\%)},
    ylabel style={font=\small},
    symbolic x coords={Java, Go, HCL, Average},
    xtick=data,
    x tick label style={font=\small},
    ymin=0, ymax=115,
    legend style={at={(0.5,-0.2)},anchor=north,legend columns=3,font=\small},
    nodes near coords,
    every node near coord/.append style={font=\tiny},
]
\addplot[fill=blue!50,draw=blue!70] coordinates {(Java,9) (Go,25) (HCL,20) (Average,18)};
\addplot[fill=green!50,draw=green!70] coordinates {(Java,27) (Go,38) (HCL,53) (Average,39)};
\addplot[fill=purple!30,draw=purple!50] coordinates {(Java,100) (Go,100) (HCL,100) (Average,100)};
\legend{Strict, Partial+, Framework Structure}
\end{axis}
\end{tikzpicture}
\caption{Cross-model agreement. 100\% structural adherence (both models follow the 4-lens protocol correctly) with 39\% finding-level agreement. The framework provides real behavioral constraint without fully determining output.}
\label{fig:crossmodel}
\end{figure}
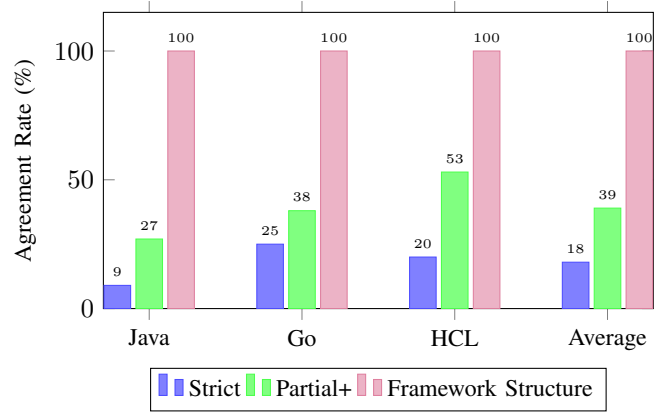

\subsection{Analysis}

Three key patterns emerge from the cross-model comparison:

\textbf{1. Framework provides structure, not determinism.} Both models consistently follow the 4-lens protocol: Cynic questions waste, Skeptic questions confidence, Ny\=aya audits logic, Confucian checks naming. Categorical structure is preserved at 100\%.

\textbf{2. Models have different attention profiles.} GPT consistently produces more adversarial, fault-finding output (average 7.0 unique findings per PR). Claude produces more balanced analysis---validating correct design AND identifying faults (average 4.3 unique findings). Both approaches are valid.

\textbf{3. Zero false positives from both models.} The framework constrains both architectures to defensible findings. The apophatic design (refusals) prevents over-extension regardless of the underlying model.

We emphasize that N=3 is insufficient for statistical generalization. These results are \emph{descriptive}, not confirmatory; they illustrate the pattern of framework-level agreement coexisting with finding-level divergence. A larger cross-model study is needed to establish whether the 39\% partial agreement rate is stable. With that caveat, the pattern is consistent with the framework providing real behavioral constraint (well above random) without fully determining output (well below ceiling).

\section{Discussion}

\subsection{What Works}

The most important result is that the framework genuinely directs attention to different concerns than generic prompting. When 51\% of findings are absent from the generic baseline output (same model, same diff, different prompt), something in the disposition structure is doing real work. The disposition-unique findings cluster in specific categories---structural questioning, operational reality checks, logical chain audits---that generic review simply does not reach. This is not ``more review''; it is a qualitatively different type of analysis.

We note that no findings were judged false-positive by the author, though this assessment was not independently validated. We did not systematically assess the false-positive rate of the generic baseline condition, which limits our ability to claim that dispositions produce \emph{fewer} false positives than unconstrained generation. The 0\% FP claim is a preliminary observation, not a validated comparative result, and confirmation bias in the author's judgment cannot be ruled out without a second rater. If the result holds under independent rating, candidate explanations include the apophatic architecture (refusals constraining output to defensible claims) and the hamartia self-check (dispositions reducing volume when over-extending), but we cannot attribute causation without controlled ablation.

The per-disposition profiles are genuinely differentiated (Table~\ref{tab:perdispo}), which validates the multi-lens architecture over a single ``thorough reviewer'' approach. Each lens accesses a distinct analytical dimension. It is worth noting that some overlap exists between Cynic and Confucian findings in naming-adjacent concerns; the distinction is not always crisp.

Perhaps the most practically useful finding is the inverse relationship with review depth (Figure~\ref{fig:depth}). The system contributes the most unique findings on under-reviewed PRs. However, the 50\% miss rate means it cannot serve as a standalone review mechanism; it is most accurately positioned as a complementary analytical layer that catches qualitatively different issues (structural, logical, relational) from those typically flagged by human reviewers.

\subsection{What Doesn't Work}

\textbf{Framework conventions (34\% of misses).} Internal framework rules (Temporal sandbox import conventions, framework-specific patterns) require codebase-specific knowledge not available in the diff. This is addressable via \emph{disposition specialization}---injecting repository context into the disposition prompt---but this has not yet been explored.

\textbf{Refactoring suggestions (22\% of misses).} Dispositions identify \emph{problems} but do not suggest \emph{restructuring}. ``This function does too much'' is a Cynic finding; ``move lines 40--60 to a helper called X'' is a refactoring suggestion that dispositions are not designed to produce.

\subsection{The Self-Debugging Property}

Each disposition's hamartia provides an intrinsic quality signal. A Cynic disposition that generates more than seven findings on a 30-line diff triggers its own self-check: ``Am I subtracting or demolishing?'' We observed this correction mechanism active in several reviews, where initial over-generation was filtered to a smaller, higher-confidence subset. The absence of author-identified false positives in our dataset is consistent with this mechanism working, though we cannot definitively attribute the precision to hamartia without a controlled ablation (dispositions with and without the self-check).

\subsection{Implications for Practice}

\textbf{Deployment model:} The system is most valuable as an additional review layer on PRs with light human review---catching structural, operational, and logical issues that busy reviewers skip. It should not replace human review but augment it.

\textbf{Per-disposition selection:} Organizations could deploy specific dispositions based on their gap profile. Teams that frequently miss naming issues deploy Confucian; teams with logic-heavy code deploy Ny\=aya; teams shipping fast with light review deploy the full Reviewer role.

\textbf{The vocabulary IS the intervention:} Naming a finding ``Ny\=aya: ungrounded inference'' rather than ``potential issue'' triggers different cognitive processing in the human reader. The philosophical vocabulary provides a taxonomy for different \emph{types} of problems, making review feedback more actionable and categorizable.

\section{Threats to Validity}

\subsection{Internal Validity}

\textbf{Training data contamination.} The model (Claude Opus) may have seen public repositories (kubernetes, envoy, elasticsearch) in training data. However: (a) internal enterprise repos (29 of 50 PRs) are not in public training data; (b) results are consistent across internal and public repos; (c) our claim is about \emph{framework direction of attention}, not novel knowledge discovery---even if the model ``knows'' the code, the framework's value is in structured attention.

\textbf{Self-validation in matching.} This is the most serious internal threat. The comparison between disposition findings and human findings was performed by the author alone; no independent rater participated. Inter-rater reliability (Cohen's kappa) has not been assessed. Given that ``same concern'' judgments require interpretation, an independent rater might disagree on 10--20\% of classifications~\cite{mcaleese2024}, shifting convergence rates by several percentage points. The false-positive rate of 0\% is particularly vulnerable to this threat: findings the author judged ``valid but debatable'' might be classified differently by another rater. We mitigate partially through: (a) explicit matching criteria requiring same code entity AND same concern type, applied consistently across all 50 PRs; (b) conservative classification where ambiguous cases default to ``unique'' rather than ``convergence''; and (c) the structured comparison protocol described in Section~IV-A. Established annotation methodologies~\cite{artstein2008} recommend independent coding of a representative subsample followed by adjudication; this remains necessary future work. We plan a 10-PR inter-rater study as the immediate next step.

\textbf{Single primary model.} All disposition reviews in Study 1 used Claude Opus. Study 2 provides preliminary cross-model evidence but is limited to N=3.

\subsection{External Validity}

\textbf{Repository selection.} Seven repositories from five organizations. Results may not generalize to other languages (Rust, JavaScript, Swift), domains (ML, frontend, embedded), or review cultures with different norms.

\textbf{Temporal scope.} PRs span 2020--2026. As AI-assisted development changes the nature of code (and code review), the framework's value proposition may shift.

\subsection{Construct Validity}

\textbf{What counts as ``same finding''?} The matching criteria require judgment. We applied conservative matching (requiring the same code entity and concern) but acknowledge this introduces subjectivity.

\textbf{Miss rate interpretation.} The raw 50.2\% miss rate overstates the analytical gap because 52\% of misses (conventions + style + typos) are categories the system intentionally does not target. The \emph{effective} miss rate for issues within disposition scope is substantially lower.

\section{Conclusion}

We have presented and empirically evaluated a system that constrains AI code review behavior through philosophical dispositions---coherent personality lenses grounded in epistemological traditions spanning 2,500 years. The key findings across 50 PRs, 7 repositories, and 5 programming languages:

\begin{enumerate}
\item The framework adds genuine value beyond generic prompting; 51\% of disposition findings are absent from the same model's generic output.
\item These unique findings are qualitatively different, targeting structural, operational, and logical concerns rather than standard code-level issues.
\item No findings were judged false-positive by the author (though independent validation is needed to confirm this).
\item Each disposition contributes a distinct analytical profile, with unique rates ranging from 71\% to 81\%.
\item Value scales inversely with human review depth, suggesting optimal deployment as a safety net for under-reviewed code.
\item Results are consistent across five languages (Python, Go, C++, Java, Terraform) and both internal and public repositories.
\item Cross-model validation confirms the framework provides structural constraint (100\% categorical adherence, 39\% finding-level agreement between models).
\end{enumerate}

The contribution is not any single disposition. Rather, it is the architectural insight that \emph{principled character}---drawn from philosophical traditions with millennia of stress-testing---can serve as a behavioral constraint mechanism occupying the space between the brittleness of fixed rules and the unpredictability of unconstrained generation. Whether this insight generalizes beyond code review to other agent tasks remains an open question.

\subsection{Future Work}

The most immediate next step is inter-rater validation: having a second researcher independently classify a 10-PR subsample to compute Cohen's kappa on the convergence/unique/miss judgments. Full cross-model validation (10+ PRs $\times$ 3 model architectures) would strengthen the framework-vs-model separation claim. Longer-term work includes longitudinal deployment in a live review pipeline with hit-rate tracking, disposition specialization with injected codebase context, and additional philosophical traditions (Aboriginal Australian epistemology, Yoruba Ifa logic) that may contribute lenses not currently represented.

\section*{Acknowledgment}

The author thanks the reviewers of the internal enterprise repositories whose review comments provided the ground truth for this study.

\smallskip\noindent\textbf{Author Role Disclosure:} The author is a member of the engineering organization that produced the internal PRs used as ground truth. The author was not among the original reviewers on any of the 29 internal PRs in the dataset; ground-truth comments are from other team members. The disposition system was designed and implemented by the author.

\smallskip\noindent\textbf{AI Assistance Disclosure:} Drafts of this manuscript were prepared with the assistance of large language models. All study design, dataset construction, empirical analysis, and conclusions are the author's own. The author reviewed and verified all generated text and is responsible for the final content of the paper.

\appendix
\section{Prompts Used}
\label{sec:prompts}

\subsection{Disposition Review Prompt}

The following prompt is the complete input to the model for the disposition condition. The model receives ONLY this prompt and the diff; no PR metadata, title, or description is provided.

\begin{lstlisting}[language={},basicstyle=\ttfamily\tiny]
You are performing a structured code review using
4 philosophical disposition lenses. Each lens asks
a different type of question. Apply ALL 4 lenses
independently to the diff below.

For each lens, produce 2-5 specific findings. Each
finding must reference specific code (file, line,
function). Do NOT produce generic advice.

### LENS 1: CYNIC (Ruthless Subtractor)
Ask: "What is hollow? What does not earn its
existence? What can be removed?"
- Refuse to accept "best practice" without
  independent justification
- Refuse to recommend addition before attempting
  subtraction

### LENS 2: SKEPTIC (Calibration Engine)
Ask: "How confident are we? What claims are
unverified? Do stakes match evidence?"
- Assign confidence levels
- Refuse to endorse without specifying credibility

### LENS 3: NYAYA (Logic Auditor)
Ask: "Is the reasoning chain valid? Are there
fallacious inferences?"
- Trace: if X changed, does Y still work?
- Refuse to pass unverified inferential steps

### LENS 4: CONFUCIAN (Naming & Relations)
Ask: "Do names match reality? What is the
relational impact?"
- Check: does renaming break callers?
- Refuse to let mismatched names persist

## Self-Check (Hamartia)
If any lens produces >7 findings on <300 lines:
- Cynic: "Am I subtracting or demolishing?"
- Skeptic: "Am I calibrating or paralyzing?"
- Nyaya: "Am I auditing or obstructing?"
- Confucian: "Am I correcting or pedanting?"
If over-extending, keep only top 3-4 findings.

[DIFF INSERTED HERE]
\end{lstlisting}

\subsection{Generic Baseline Prompt}

\begin{lstlisting}[language={},basicstyle=\ttfamily\tiny]
You are an expert code reviewer. Review this diff
thoroughly. List all issues found. Be specific
about file and line numbers.

[DIFF INSERTED HERE]
\end{lstlisting}

Both conditions use the same model (Claude Opus, Anthropic), default temperature, and receive only the diff as context.

\balance
\bibliographystyle{IEEEtran}

\end{document}